\documentstyle[epsf,prl,aps,preprint]{revtex}
\draft
\begin{document} 
%@twocolumnfalse\endcsname
\title{\textbf{Geometric Formulation of Unique Quantum Stress Fields}} 
\author{Christopher L. Rogers and Andrew M. Rappe}
\address{Department of Chemistry and Laboratory for Research 
on the Structure of Matter, \\
University of Pennsylvania, Philadelphia, PA 19104-6323.}  
\date{\today}
\maketitle

\begin{abstract}
We present a derivation of the stress field for an interacting
quantum system within the framework of local density functional theory.
The formulation is geometric in nature and exploits the relationship
between the strain tensor field and Riemannian metric tensor field.
The resultant expression obtained for the stress field is gauge-invariant 
with respect to choice of energy density, and therefore
provides a unique, well-defined quantity. To illustrate this
formalism, we compute the pressure field for two phases of solid
molecular hydrogen.

\noindent
\end{abstract}
%]
The stress, or the energetic response to deformation or strain, plays
an important role in linking the physical properties of a material
(e.g.\ strength, toughness) with the behavior of its microstructure.
In addition, the spatial distribution of stress is an invaluable tool
for continuum modeling of the response of materials. The stress
concept has been applied at atomistic scales as well. Over the last
fifteen years, there has been a continuing trend toward understanding
various structural and quantum-mechanical phenomena in materials in
terms of their response to stress \cite{ibachrev}.

For example, the residual stress at equilibrium has been used to
assess the structural stability of systems containing surfaces or
strained interfaces. It has been demonstrated that the
desire to minimize surface stress can give rise to reconstructions on
high symmetry surfaces
\cite{needs1,needs2,scheffler,ibach,filippeti1,vanderbilt1,vanderbilt2},
and the stability of epitaxially grown bimetallic systems has been
attributed to the formation of incommensurate overlayers, defects, and
dislocations which minimize the stress near the metal-metal interface
\cite{agcu1,agcu2}. The stress can have significant effects on
chemical reactivity as well. It has been shown that small molecule
chemisorption energies and reaction barriers on certain strained metal
and strained semiconductor surfaces are quite different from those on
the unstrained surface \cite{vanderbilt3,norskov}.

Formally, studies of the above phenomena must include a
quantum-mechanical description of the system's electronic degrees of
freedom. Therefore, one must consider how a stress is defined quantum
mechanically. Methods for calculating the stress in quantum mechanical
systems have been developed since the birth of quantum theory itself
\cite{schro}. However, research in developing formalisms for
determining the quantum stress in solid-state systems has recently
been revitalized. This is mainly due to ever-increasing opportunities
to perform accurate and efficient quantum-mechanical calculations on
systems which exhibit stress mediated phenomena.

The stress is a rank-two tensor quantity, usually taken to be
symmetric and therefore torque-free. Two useful representations of the
stress tensor are the volume-averaged or total stress, $T_{\alpha
\beta}$, and the spatially varying stress field $\sigma_{\alpha
\beta}(\bbox{\mathrm{x}})$. The two representations are related since
the total stress for a particular region in a system is the stress
field integrated over the volume. Nielsen and Martin developed a
formalism for calculating the total quantum stress in periodic systems
\cite{nielsen}. They define the total stress as the variation of the
total ground-state energy with respect to a uniform scaling of the
entire system. This uniform scaling corresponds to a homogeneous or
averaged strain over the entire system. They further demonstrate that
the total quantum stress is a unique and well-defined physical
quantity. Their formulation has been successfully implemented to
study a variety of solid state systems \cite{needs1,needs2}. Other
formalisms for determining the total quantum stress have been created
as well \cite{needs1,passerone,resta}.

Although these formalisms have provided important tools for studying
quantum stress, the stress field is a more useful quantity that
contains important information regarding the distribution of the
stress throughout the system. A knowledge of the spatial dependence
of the quantum stress is vital if one wishes to predict the spatial
extent of structural modifications or understand phenomena at
interfaces in complex heterogeneous systems. However, certain
definitions of the quantum stress field suggest that it can only
be specified up to a gauge. (This ambiguity manifests itself
in classical atomistic models as well.) It has therefore been
asserted that the quantum stress field is not a well-defined physical
quantity, even though physical intuition may suggest otherwise. A
traditional way to develop a quantum stress field formalism is to
consider the stress field's relationship with the force field. From
this perspective, the stress field can be defined as any rank-two
tensor field whose divergence is the force field of the system:
\begin{eqnarray}
F^{\alpha}= \nabla_{\beta} \sigma^{\alpha
\beta}. \label{force-stress}
\end{eqnarray}
(Note that the Einstein summation convention for repeated indices is used throughout the Letter.)
One can add to $\sigma^{\alpha \beta}$ a gauge of the form
\begin{eqnarray}
\frac{\partial }{\partial x^{\gamma}}A^{\alpha \beta \gamma}
(\bbox{\mathrm{x}}),
\end{eqnarray}
where $A^{\alpha \beta \gamma}$ is any tensor field antisymmetric in
$\beta$ and $\gamma$, and recover the same force field, thereby
demonstrating the non-uniqueness of this stress field definition. General
formulations for computing non-gauge-invariant stress fields in
quantum many-body systems have been derived by Nielsen and Martin,
Folland, Ziesche and co-workers, and Godfrey
\cite{nielsen,folland,ziesche,godfrey}. There have been several
attempts to overcome this problem of non-uniqueness. For example, the
stress field formalism of Chen and co-workers has been applied to
numerous solid state systems to determine the local pressure around a
region \cite{chen}. However, their method assumes that the potential is
pair-wise only. Several \textit{ab-initio} quantum stress field
formulations have been developed, as well. Ramer and co-workers
developed a method to calculate the resultant stress field from an
induced homogeneous strain \cite{ramer}. They incorporate the
additional constraint that the field must be the smoothest fit to the
ionic forces. This method cannot be used to calculate the residual
stress field at equilibrium, nor can it determine the energy
dependence on strains which do not have the periodicity of the unit
cell. Filippetti and Fiorentini developed a formulation of the stress
field based on the energy density formalism of Chetty and Martin
\cite{filippetti2,chetty}. Since this formulation is based explicitly
on the energy density, which is not gauge-invariant, the resultant
stress field is not unique. Mistura succeeded in developing a general
gauge-invariant formalism for pressure tensor fields of inhomogeneous
fluids within classical statistical models using a Riemannian
geometric approach \cite{mistura}.

This Letter extends Mistura's work, developing a Riemannian geometric
formalism for computing gauge-invariant stress fields in quantum
systems within the local density approximation (LDA) of density
functional theory (DFT). We show that the response of the total
ground-state energy of a quantum system to a local spatially varying
strain is a unique and physically meaningful field quantity which can
be determined at every point in the system.

Using a procedure well known in continuum theory (see for example
Ref.\ \cite{fung}), one can formally relate a Riemannian metric tensor
field $g_{\alpha \beta}(\bbox{\mathrm{x}})$ with the strain field
$\epsilon_{\alpha \beta}(\bbox{\mathrm{x}})$:
$g_{\alpha \beta}= \delta_{\alpha \beta} + 2\epsilon_{\alpha \beta}$,
with
\begin{equation}
\epsilon_{\alpha \beta} \equiv \frac{1}{2} \ (\delta_{\alpha \gamma}
\partial_{\beta} u^{\gamma} + \delta_{\gamma
\beta}\partial_{\alpha}u^{\gamma} + \delta_{\gamma
\kappa}\partial_{\alpha} u^{\gamma} \partial_{\beta}u^{\kappa}),
\label{strain}
\end{equation} 
where $\partial_\alpha \equiv \partial/\partial x^{\alpha}$ \cite{landau}.  Here the
strain field is defined in terms of a vector displacement field
$u^{\alpha}$ which maps coordinates $x^{\alpha}$ in the non-deformed
system, to the coordinates $x^{'\alpha}=x^{\alpha}+u^{\alpha}$ in the
deformed system.

The stress field $\sigma^{\alpha \beta}(\bbox{\mathrm{x}})$
and strain field obey a virtual work theorem expressing the
energy response to variations in the strain:
\begin{equation}
\delta E= \int \sqrt{g} \sigma^{\alpha \beta} \delta \epsilon_{\alpha
\beta} d^3 x, \label{vari-E}
\end{equation}
where $g(\bbox{\mathrm{x}})$ is the determinant of 
$g_{\alpha\beta}(\bbox{\mathrm{x}})$.  
It can be shown that the stress field is related to the functional 
derivative of the energy with respect to the metric field \cite{mistura}:
\begin{eqnarray}
\sigma^{\alpha \beta} &\equiv& \frac{1}{\sqrt{g}} \frac{\delta E}{\delta \epsilon_{\alpha \beta}} = \frac{2}{\sqrt{g}} \frac{ \delta E}{\delta g_{\alpha \beta}}, \nonumber\\ 
\sigma_{\alpha\beta}&=&-\frac{2}{\sqrt{g}} \frac{ \delta E}{\delta g^{\alpha\beta}}.\label{stress-metric}
\end{eqnarray}

We now derive the quantum stress field of a many-electron system in
the presence of a fixed set of classical positive charged ions using
local density functional theory \cite{HK,KS}.  The ground state
electronic charge density of the system is written as
$n(\bbox{\mathrm{x}})= \sum_i
\phi^{\ast}_{i}(\bbox{\mathrm{x}})\phi_{i}(\bbox{\mathrm{x}})$, where
$\phi_i$ are single-particle orthonormal wavefunctions. 
For this derivation, we assume orbitals with fixed integer occupation
numbers. The extension to metals with Fermi fillings is straightforward, 
simply necessitating use of the Mermin functional instead of the total energy \cite{mermin}.
The total charge density of the system can be written as a sum over all
ionic charges and $n$:
\begin{eqnarray}
\rho(\bbox{\mathrm{x}}) = \sum_i 
\frac{Z_i}{\sqrt{g}}\delta(\bbox{\mathrm{x}} - \bbox{\mathrm{R}}_i) -
n(\bbox{\mathrm{x}}),
\end{eqnarray}
where $Z_i$ is the charge of the $i$-th ion located at position
$\bbox{\mathrm{R}}_i$, and the presence of $\sqrt{g}$ insures proper
normalization of the delta function.  The energy of the system can be
written as the following constrained functional:
\begin{equation}
E = E_{\mathrm{k}} + E_{\mathrm{Coulomb}} +
E_{\mathrm{xc}} - \sum_{i} \lambda_{i} \left( \int \sqrt{g}
\phi^{\ast}_{i} \phi_{i} d^3x - 1 \right). \label{totale}
\end{equation}
Here $E_{\mathrm{k}}$ is the single particle kinetic energy,
$E_{\mathrm{Coulomb}}$ is the classical Coulomb interaction between
the total charge density and itself, and $E_{\mathrm{xc}}$ is the
exchange-correlation energy of the electrons. The appearance of the
last term in Eq.\ \ref{totale} is due to the orthonormality constraint
of the orbitals. (We choose a unitary transformation on $\left
\{\phi_{i} \right \}$ which enforces orthogonality.) One can express
$E$ as an integral over an energy density \cite{chetty}. The choice
of energy density gauge will not affect the derived stress field,
since all our results depend only on the total energy $E$.  
For convenience, we express the energy
terms in Eq.\ \ref{totale} as the following:
\begin{eqnarray}
E_{\mathrm{k}} &=& \frac{1}{2} \int \sqrt{g} \sum_i g^{\alpha \beta}
\partial_{\alpha} \phi^{\ast}_{i} \partial_{\beta} \phi_{i} d^3x
\nonumber \\ E_{\mathrm{Coulomb}} &=& \int \sqrt{g} \left( \rho V -
\frac{1}{8\pi} g^{\alpha \beta} {\mathcal{F}}_{\alpha}
{\mathcal{F}}_{\beta} \right) d^3x \nonumber \\ E_{\mathrm{xc}} &=&
\int \sqrt{g} \ n \varepsilon_{\mathrm{LDA}}(n) d^3x,
\end{eqnarray}
where ${\mathcal{F}}_{\alpha} = - \partial_{\alpha}V $ is the electric
field due to the Coulomb potential $V$ generated by $\rho$, and
$\varepsilon_{\mathrm{LDA}}(n)$ is the LDA exchange-correlation
energy density.
To obtain the electronic ground-state energy, we require $\delta E /
\delta \phi^{\ast}_i = 0$ with the additional constraints of a fixed metric
($\delta g_{\alpha \beta} =0$) and a fixed ionic charge density
($\delta \rho = -\delta n $). This implies that the orbitals must obey
the Euler-Lagrange equations
\begin{equation}
-\frac{1}{2\sqrt{g}} \partial_{\alpha} \left(\sqrt{g} g^{\alpha \beta}
\partial_{\beta} \phi_i \right) + \frac{\delta
E_{\mathrm{Coulomb}}}{\delta n} \phi_i + \frac{\delta
E_{\mathrm{xc}}}{\delta n} \phi_i = \lambda_i \phi_i, \label{E-L}
\end{equation}
which can be considered the Kohn-Sham equations in curvilinear
coordinates. Also, a least-action principle for $E_{\mathrm{Co\vspace{0.5cm}ulomb}}$ 
requires that $\rho$ and $V$ obey the Poisson equation:
\begin{equation}
\frac{1}{\sqrt{g}} \partial_{\alpha}\left(\sqrt{g} g^{\alpha \beta}
\partial_{\beta}V \right) = -4\pi \rho. \label{Poisson}
\end{equation}

We now vary the total energy with respect to the
metric. It can be proven that we do not need to consider
variations in the electronic wavefunctions, charge density and potentials, since all
such variations would vanish due to Eq.\ \ref{E-L} and Eq.\
\ref{Poisson}. This is the same principle used in the derivation of
the Hellmann-Feynman force theorem and the
energy-momentum tensor (the variation of the action with respect to metric) in general relativity
\cite{hellmann,feynman,landau2}. If a different electromagnetic gauge is
chosen, variations with respect to the potential $V$ would have to be
computed explicitly; this change has no impact on the stress field in
Eq.\ \ref{stressfield}.   
%Therefore, for a general energy
%density ${\mathcal{E}}$ the stress field is:
%\begin{eqnarray}
%\sigma_{\alpha \beta} &=& -\frac{2}{\sqrt{g}} \left \{
%\frac{\partial}{\partial g^{\alpha \beta}}\left[ \sqrt{g}{\mathcal{E}} 
%- \sqrt{g}\sum_{i} \lambda_{i} \phi^{\ast}_{i}\phi_{i} \right] \nonumber \right. \\
%& & \left. \mbox{} - \partial_{\gamma} \left( \frac{\partial }{\partial \left(
%\partial_{\gamma} g^{\alpha \beta} \right)} \left[
%\sqrt{g}{\mathcal{E}} - \sqrt{g}\sum_{i} \lambda_{i}
%\phi^{\ast}_{i}\phi_{i} \right] \right)\right \}. \label{maineq}
%\end{eqnarray} 
%Evaluating Eq.\ \ref{maineq} for local density functional theory yields
Performing the variation of the total energy with respect to the metric gives the stress field in local density functional theory as:
\begin{eqnarray}
\sigma_{\alpha \beta} &=& - \sum_i \partial_{\alpha}
\phi^{\ast}_{i} \partial_{\beta} \phi_{i} + \frac{1}{4\pi}
{\mathcal{F}}_{\alpha}{\mathcal{F}}_{\beta} + g_{\alpha \beta} \left(
\frac{1}{2} \sum_i \partial_{\gamma}\phi^{\ast}_{i} \partial^{\gamma}\phi_{i} \nonumber \right.  \\
& & \left. \mbox{} -\frac{1}{8 \pi} {\mathcal{F}}_{\gamma} {\mathcal{F}}^{\gamma} + n
\varepsilon_{\mathrm{LDA}}(n) - \sum_i \phi^{\ast}_{i} \phi_{i} \left[
\lambda_i + V \right] \right) , \label{stressfield}
\end{eqnarray}
where we have used the relation $ \partial \sqrt{g} /\partial
g^{\alpha \beta} = -\frac{1}{2}\sqrt{g}g_{\alpha \beta}$.  
Using Eq.\ \ref{E-L}, we can rewrite Eq.\ \ref{stressfield} as
\begin{eqnarray}
\sigma_{\alpha \beta} &=& - \sum_i \partial_{\alpha}
\phi^{\ast}_{i} \partial_{\beta} \phi_{i} + \frac{1}{4\pi}
{\mathcal{F}}_{\alpha}{\mathcal{F}}_{\beta} + g_{\alpha \beta} \left(
\frac{1}{2} \sum_i \partial_{\gamma}\phi^{\ast}_{i} \partial^{\gamma}\phi_{i} \nonumber \right.  \\
& & \left. \mbox{} + \frac{1}{2\sqrt{g}} \sum_i \phi^{\ast}_{i} \partial_{\kappa}
\left( \sqrt{g} g^{\kappa \gamma} \partial_{\gamma} \phi_{i} \right)  
- \frac{1}{8 \pi} {\mathcal{F}}_{\gamma} {\mathcal{F}}^{\gamma} + n \left(
\varepsilon_{\mathrm{LDA}}(n) - \frac{\delta E_{\mathrm{xc}}}{\delta n} \right) 
\right) . \label{stressfield2}
%\phi^{\ast}_{i} \phi_{i} \left[
%\lambda_i + V \right] \right) \right \},  \label{stressfield}
\end{eqnarray}
When Eq.\ \ref{stressfield2} is evaluated at the Euclidean metric 
($g_{\alpha \beta} = \delta_{\alpha \beta}$), it gives the stress
field at zero applied strain, and the $\left \{\phi_i \right \}$ are then
solutions to the standard Kohn-Sham equations. From here on, we
will refer to $\sigma_{\alpha \beta}$ with an implied evaluation at
the Euclidean metric.

It is important to note several key features of the form of
$\sigma_{\alpha \beta}$. First, the Coulombic contribution to the
quantum stress field is equivalent to the classical Maxwell stress
field. This Coulombic term can be obtained by Filippetti and Fiorentini's
formalism in Ref.\ \cite{filippetti2} if one chooses the Maxwell
gauge. Also, the contribution of the exchange-correlation energy to
our stress field is only in the diagonal (pressure-like) terms, which
is the proper behavior for local density functionals \cite{nielsen}
and is identical to the exchange-correlation stress derived in Ref.\
\cite{filippetti2}.  However, the kinetic contribution to 
Eq.\ \ref{stressfield2} is unique to our derivation. 
Our stress field contains diagonal terms which are
similar to the symmetric and antisymmetric kinetic energy densities.
%First, there is a strong similarity between
%the classical Maxwell stress field and the quantum stress
%field. However, the presence of the eigenvalue $\lambda_{i}$ in the
%diagonal (pressure) terms of the quantum stress field represents a
%fundamental coupling between kinetic-like and potential-like terms.  Also,
%the contribution of the exchange-correlation energy to the stress
%field is only in the diagonal terms, which is the proper behavior for
%local density functionals \cite{nielsen}. 
By integrating the stress
field over all space, we can obtain the total stress $T_{\alpha
\beta}$:
\begin{eqnarray}
T_{\alpha \beta} &=& \int \left \{ - \sum_{i} \partial_{\alpha}
\phi^{\ast}_{i} \partial_{\beta} \phi_{i} + \frac{1}{4\pi}
{\mathcal{F}}_{\alpha}{\mathcal{F}}_{\beta} - \delta_{\alpha \beta}
\frac{1}{8\pi} {\mathcal{F}}_{\gamma}{\mathcal{F}}^{\gamma} \nonumber \right. \\
& & \left. \mbox{} + \delta_{\alpha \beta} n \left(\varepsilon_{\mathrm{LDA}}(n) -
\frac{\delta E_{\mathrm{xc}}}{\delta n} \right) \right \} d^3x,
\end{eqnarray} 
which is identical to the expression derived by Nielsen and
Martin\cite{nielsen}. 

In order to demonstrate the utility of our stress field formalism,
we compute within DFT the pressure field, $\frac{1}{3}\left(\sigma_{11} + \sigma_{22}
+ \sigma_{33} \right)$, for two phases of solid molecular hydrogen 
under external hydrostatic pressure of 50 GPa. \cite{calc1}.
Both structures consist of stacked two-dimensional triangular
lattices of hydrogen molecules, with the molecular axis parallel to
the stacking direction and a repeat unit of two layers.  The \textit{m}-hcp
structure has alternating layers shifted so that each hydrogen
molecule is directly above triangular hollow sites in the neighboring
layers.  The second structure (belonging to the \textit{Cmca} space group) has
a different shift, so that each molecule lies directly above midpoints
between nearest-neighbor pairs of molecules in adjacent layers \cite{solidh1}.
The energetics and electronic properties of both structures have been studied 
extensively from first principles \cite{solidh2,solidh3}.

Examination of the pressure field permits us to rationalize the energy
ordering of these structures.  The \textit{m}-hcp structure is
energetically favored by 60 meV/molecule.  Figure \ref{fig1}A shows a
contour plot of the pressure field for the \textit{Cmca}
structure. The pressure is tensile (greater than zero) through the
interstitial region, indicating that contraction is locally favorable.
The pressure is greatest in the volume directly above and below each
molecule, averaging 3 eV/\AA$^3$. This implies that the system would
energetically favor increased intermolecular coordination. In Figure
\ref{fig1}B we show a similar plot for the \textit{m}-hcp
structure. Again the pressure within the interstitial region is
tensile. However, the pressure field has significantly rearranged, and
the pressures in the regions above and below molecules have been
reduced to approximately 2.25 eV/\AA$^3$.  It is also clear from the
charge density plots (Figure \ref{fig1}C and \ref{fig1}D) that the
reduction in pressure is correlated with an increase in bonding
between molecules and increased charge delocalization.  The pressure
fields within the molecules are compressive, and they are several
orders of magnitude larger than the interstitial features.  However,
because these large fields are very similar in both structural phases
(and the free molecule), they are not important for understanding
relative phase stability. Thus, changes in the pressure stress field
highlight regions and charge density features that contribute to
favorable energetic changes.

We have developed a formulation for unambiguously
determining the stress field in an interacting quantum system
described by local density functional theory.  The resultant
expression for the stress field is gauge-invariant with respect to
choice of energy density and can be obtained via a method analogous
to the computation of Hellmann-Feynman forces. 
Our application of this formalism to solid molecular
hydrogen demonstrates that a stress field analysis can associate
energetics with particular microscopic structural features of a
material. This stress field formulation has the potential to
become an invaluable aid to the understanding of structural phenomena
in complex solid-state systems.

\section*{Acknowledgements}
The authors wish to acknowledge R.\ M.\ Martin, E.\ J.\ Mele,  
P.\ Nelson, M.\ Scheffler, and D.\ Vanderbilt for their comments and valuable
suggestions regarding this work, and E.\ J.\ Walter for his assistance
with the numerical calculations.  This work was supported by NSF grant
DMR 97-02514 and the Air Force Office of Scientific Research, Air
Force Materiel Command, USAF, under grant number F49620-00-1-0170. AMR
would like to thank the Camille and Henry Dreyfus Foundation for
support. Computational support was provided by the National Center for
Supercomputing Applications and the San Diego Supercomputer Center.

\begin{figure}
\epsfysize=6.5in
\centerline{}

\vspace{0.5cm}
\caption{
Contour plots of the pressure field and charge density within DFT for
the \textit{Cmca} structure (panels A and C) and for the \textit{m}-hcp structure (B and
D) of solid hydrogen.  The vertical axis is the stacking direction for
the layers, and the horizontal axis is the direction along which
alternating layers are shifted.  The plots are 6.794 \AA ~high and 7.206 \AA 
~wide.  Ten contours are shown, over a range of 0--3~eV/\AA$^3$ for the
pressure fields, and from 0--2.5~\textit{e}/\AA$^3$ for the charge densities.
}
\label{fig1}
\end{figure}


\begin{references}
\bibitem{calc2} Calculations for both structures were performed in an orthorhombic unit cell with parameters $a=2.08$, $b$, and $c$ with $b/a=\sqrt{3}$ and
$c/a=\sqrt{8/3}$. All atomic forces were relaxed to $<$ 5 meV/\AA.
\end{references}
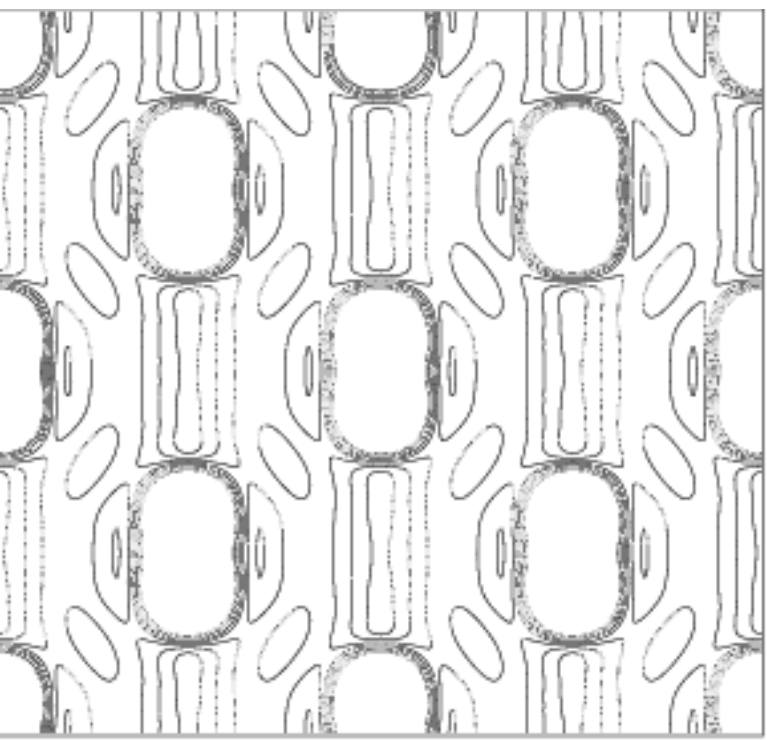
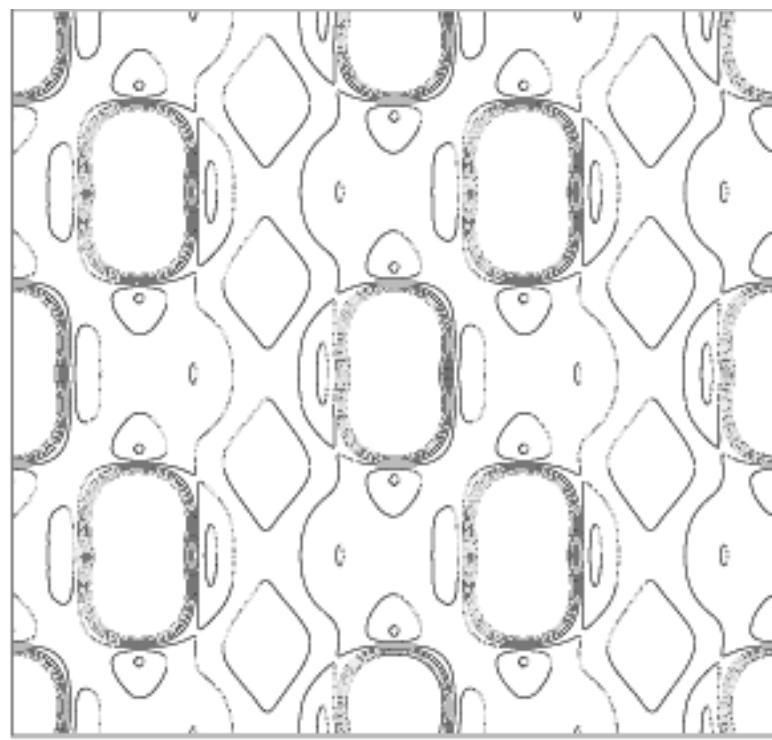
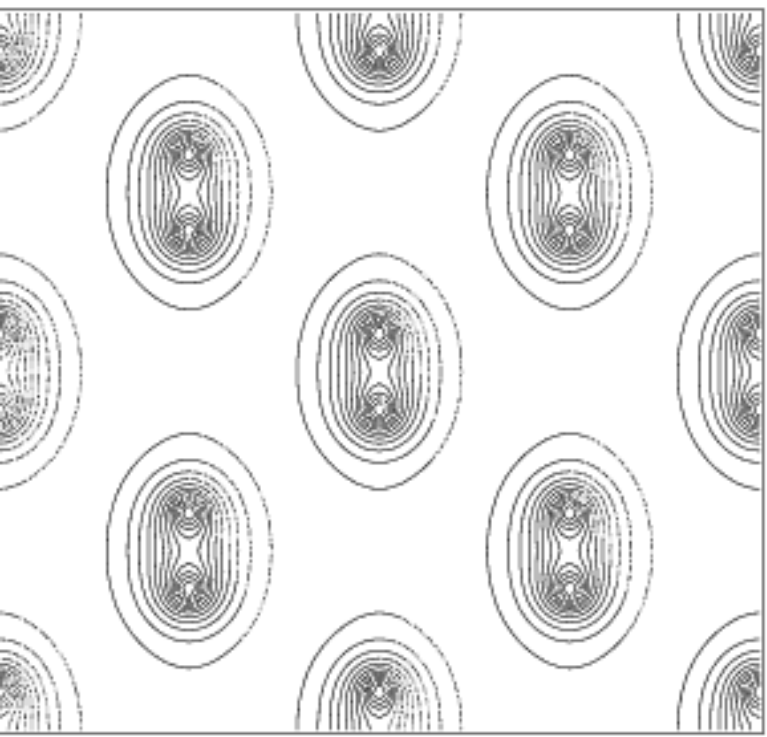
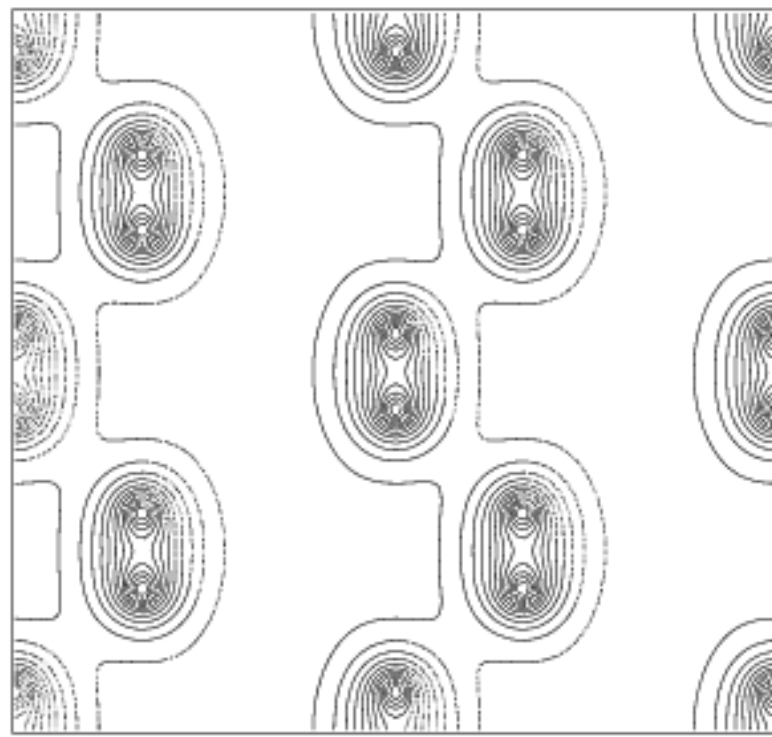
\end{document}